\documentclass{aa}
\usepackage{txfonts}
\usepackage{graphicx}
\usepackage[]{natbib}

\begin{document}

\title{Environments of galaxies in groups within the supercluster-void network}

\author{H. Lietzen \inst{1}
  \and E. Tempel \inst{2,3}
  \and P. Hein\"am\"aki \inst{1} 
  \and P. Nurmi \inst{1}
  \and M. Einasto \inst{3}
  \and E. Saar \inst{3}}

\institute{Tuorla Observatory, Department of Physics and Astronomy, 
 University of Turku, V\"ais\"al\"antie 20, 21500 Piikki\"o, Finland
 \and National Institute of Chemical Physics and Biophysics, Tallinn 10143, Estonia
 \and Tartu Observatory, 61602 T\~oravere, Tartumaa, Estonia}

\date{Received / Accepted }

\abstract
{The majority of all galaxies reside in groups of fewer than 50 member galaxies. These groups are distributed in various large-scale environments from voids to superclusters. }
{The evolution of galaxies is affected by the environment in which they reside. Our aim is to study the effects of the local group scale and the supercluster scale environments on galaxy evolution. }
{ {We use a luminosity-density field to determine the density of the large-scale environment of galaxies in groups of various richnesses.  We calculate the fractions of different types of galaxies in groups with richnesses of up to 50 member galaxies and in different large-scale environments from voids to superclusters. }
}
{The fraction of passive elliptical galaxies rises and the fraction of star-forming spiral galaxies declines when the richness of a group of galaxies rises from two to approximately ten galaxies.  On large scales, passive elliptical galaxies become more numerous than star-forming spirals when the environmental density grows to values typical of superclusters. The large-scale environment affects the level of these fractions in groups: galaxies in equally rich groups are more likely to be elliptical in supercluster environments than at lower densities. {The crossing point, where the number of passive and star-forming galaxies is equal, occurs in superclusters in groups that are of lower richness than in voids. Galaxies in low-density environments need to occupy richer groups to evolve from star-forming to passive than galaxies in high-density environments.}  Groups in superclusters are on average more luminous than groups in large-scale environments of lower density. These results imply that the large-scale environment affects the properties of galaxies and groups.}
{ {Our results suggest that the evolution of galaxies is affected by both, the group in which the galaxy resides and its large-scale environment. Galaxies in lower-density regions develop later than galaxies in similar mass groups in high-density environments.}}


\maketitle

\section{Introduction}

Galaxies form groups and clusters of various sizes. These groups and clusters are distributed in different environments on the large scale. Regions of high galaxy density are superclusters, which are surrounded by filaments and low-density void areas.

 {The properties of galaxies are affected by their cluster- or group-scale environment. In group environments, galaxies experience baryonic processes, such as ram pressure stripping \citep{Gunn1972}, viscous stripping, and strangulation \citep{Larson1980}, and gravitational effects, such as galaxy tidal interactions with other galaxies and the cluster potential \citep{Moore1996} and galaxy merging. These local processes are believed to modify galaxy morphologies from spiral galaxies to gas poor and spheroidal ones \citep{Gunn1972, Dressler1980, Postman1984}. }
According to the morphology-density relation, elliptical galaxies are more concentrated at the centers of clusters than spiral galaxies  {\citep{Dressler1980,Postman1984}}. Similarly, galaxies in higher density environments are more luminous \citep{Hamilton1988} and have lower star-formation rates \citep{Gomez2003}.  {The dependence of either the star-formation rate or color on the environment is stronger than the dependency between morphology and the environment \citep{Kauffmann2004,Blanton2005}. According to \citet{Baldry2006}, the environmental dependence is at least as important as stellar mass in determining the fraction of red galaxies in a population.} Finally, different types of active galactic nuclei (AGNs) appear in different local environments \citep{Hickox2009}.

In addition to the local group or cluster scale, the properties of galaxies also depend on their environment on large scales. The large-scale morphology-density relation was first found by \citet{Einasto1987}. According to \citet{Balogh2004}, the fraction of red galaxies is higher where the surface density of galaxies is higher. \citet{Porter2008} found that the star-formation rate of galaxies depends on their location in large-scale filaments. \citet{Skibba2009} found significant color-environment correlations on the 10\,$h^{-1}$Mpc scale in both spiral and elliptical galaxies. \citet{Tempel2011} derived luminosity functions of spiral and elliptical galaxies in different large-scale environments. They found that the luminosity function of elliptical galaxies depends strongly on environment, while for spiral galaxies the luminosity function is almost independent of the  {large-scale} environment. The large-scale environments of AGN were studied by \citet{Lietzen2011}. In the same way as on smaller scales, radio galaxies favor high-density environments, while radio-quiet quasars and Seyfert galaxies are mostly located in low-density regions.

In high-density large-scale environments, groups and clusters of galaxies tend to be larger and more massive than in low-density environments \citep{Einasto2005}. {According to \citet{Einasto2012a}, isolated clusters are also poorer than supercluster members.} Because of this, group-scale effects are also present when studying the large-scale environments.  {The effects on galaxy properties on different scales up to tens of Mpc  can be distinguished using a luminosity-density method \citep{Einasto2003}. Group and supercluster scale environments of galaxies were studied together by \citet{Einasto2008}.} They found that in the outskirt regions of superclusters, rich groups contain more late-type galaxies than in the supercluster cores, where the rich groups are populated mostly by early-type galaxies. \citet{Einasto2007} found that in the high-density cores of rich superclusters there is an excess of early-type galaxies in groups and clusters, as well as among galaxies that do not belong to any group.  

 {Some studies have used the number-density of galaxies to study the large-scale environments of galaxies. In these studies, the large scale usually refers to scales of a few Mpc. \citet{Zandivarez2011} found that the Schechter parameters of the luminosity functions of galaxies in groups in high-density environments do not depend on the mass of the group, while in low-density regions some dependence does occur. On the other hand, \citet{Blanton2007} found that on a few Mpc scales the environment is only weakly related to the colors of galaxies, while the small-scale environment matters more. \citet{Wilman2010} found no correlation on scales of $\sim$1\,Mpc and even an anticorrelation on scales of 2--3\,Mpc in the fraction of red galaxies. }

Of the different ways of determining the environment of a galaxy, some are sensitive to the local scales, which in turn depend on the size of the dark matter halo surrounding the galaxy. Other methods are more sensitive to large scales, which represent the environment in the supercluster-void network  {\citep{Haas2011,Muldrew2012}}. In this paper, we use spectroscopic galaxy and group catalogs based on the Sloan Digital Sky Survey (SDSS) to study the environments of galaxies in groups on both local and large scales. As a measure of the local-scale environment, we use the richness of the group, and for the large-scale environment a luminosity-density field smoothed to scales typical of superclusters. Our goal is to distinguish between the effects that different scales of environment have on galaxies.

 {The paper is composed as follows: in Sect. \ref{Data}, we present the data and describe how the group catalog and the large-scale luminosity-density field were constructed. We also describe the galaxy classification criteria. In Sect. \ref{Results}, we present our results on the environments of galaxies on both the group and large scales. In Sect. \ref{Discussion}, we compare our results to previous studies and discuss the possible implications of our results for galaxy evolution.  }

%ET: I added citation to WMAP 7-year results.
 {Throughout this paper we assume a cosmological model with a total matter density $\Omega_{\mbox{m}}=0.27$, dark energy density $\Omega_\Lambda=0.73$, and  Hubble constant $H_0=100h$\,km\,s$^{-1}$Mpc$^{-1}$ \citep{Komatsu:11}. }

\section{Data}\label{Data}

\subsection{Galaxy and group sample}

We used flux-limited galaxy and group catalogs based on the eighth data release (DR8) of the SDSS \citep{York:00, Aihara:11}.  {The details of the galaxy and group catalogs used in this study are given in \citet{Tempel:12}.} We determined groups of galaxies using the friend-of-friend cluster analysis that was introduced iton cosmology by \citet{Turner1976}. This method was named friends-of-friends (FoF) by \citet{Press1982}. With the FoF method, galaxies are linked into groups using a certain linking length. In a flux-limited sample, the density of galaxies slowly decreases with distance. To take this selection effect properly into account, we rescaled the linking length with distance, calibrating the scaling relation with observed groups \citep[see][for details]{Tago2008, Tago2010}. The linking length in our group-finding algorithm increases moderately with distance. 

 {The most important problem in obtaining the catalogs of systems of galaxies has been the inhomogeneity of the resulting samples owing to the selection effects of the various search procedures. The choice of the method and parameters depends on the goal of the study. For example, \citet{Berlind:06} applied the FoF method to the volume-limited samples of the SDSS with the goal of measuring the group multiplicity function and constraining the properties of the dark matter halos. \citet{Yang2007} applied a halo-based group finder to study the relation between galaxies and dark matter halos over a wide dynamic range of halo masses. Our goal is to use groups and clusters for large-scale structure studies. In this respect, our group catalog is rather homogeneous: the group richnesses, mean sizes, and velocity dispersions are practically independent of the group distance.}

Since the data are magnitude-limited, our study has distance-dependent selection effects: at the largest distances, only the most luminous galaxies are detected. This luminosity-dependence also causes a color and/or morphology dependence, as the red, elliptical galaxies tend to be more luminous than blue, spiral galaxies. To reduce the distance effects, we limited our data to distances between 120 and 340\,$h^{-1}$Mpc (the redshift range 0.04--0.116) for which we have the most reliable galaxy data. At distances smaller than 120\,$h^{-1}$Mpc, there are extremely rich clusters that are not seen at larger distances. At distances larger than 340\,$h^{-1}$Mpc, the richness of groups declines rapidly \citep{Tago2010}. After limiting the distance, our data contains 306\,397 galaxies belonging to 45\,922 groups. Despite this limitation, the average color of galaxies still depends on the distance. The average $g-r$ color of galaxies at distances of from 120 to 140\,$h^{-1}$Mpc is 2.3, while for galaxies from 320 to 340\,$h^{-1}$Mpc it is 2.9.

 {The apparent magnitude $m$ (corrected for Galactic extinction) was transformed into the absolute magnitude $M$ according to the usual formula}
\begin{equation}
M_\lambda = m_\lambda - 25 -5\log_{10}(d_L)-K,
\end{equation} 
 {where $d_L$ is the luminosity distance in units of $h^{-1}$Mpc, $K$ is the $k$+$e$-correction, and the index $\lambda$ refers to the $ugriz$ filters. The $k$-corrections were calculated with the \mbox{KCORRECT\,(v4\_2)} algorithm \citep{Blanton:07} and the evolution corrections were estimated, using the luminosity evolution model of \citet{Blanton:03}. The magnitudes correspond to the rest-frame (at redshift $z=0$).}

\subsection{Estimating the environmental densities}

As a measure of the large-scale environment, we used a luminosity-density field that is smoothed to scales characteristic of superclusters. To construct the luminosity-density field, the luminosities of galaxies were corrected by a weighting factor to take into account the luminosities of galaxies outside the magnitude window of the survey. The weighting factor $W_L(d)$ was defined as
\begin{equation}
W_L(d)=\frac{\int_0^\infty \! L\phi(L) \, \mathrm{d} L}{\int_{L_1(d)}^{L_2(d)} \! L\phi(L) \, \mathrm{d} L},
\end{equation} 
{where $L_{1,2}=L_\odot 10^{0.4(M_\odot - M_{1,2})}$ are the luminosity limits of the observational window at the distance $d$ corresponding to the absolute magnitude limits of the window from $M_1$ to $M_2$; we took $M_\odot=4.64$\,mag in the $r$-band \citep{Blanton:07}, and $\phi(L)$ denotes the galaxy luminosity function.} The luminosity function was calculated as described in \citet{Tempel2011}, and can be approximated by the double power-law
\begin{equation}
\phi(L)\mathrm{d}(L)\propto(L/L^*)^\gamma)^{(\delta-\alpha)/\gamma}\mathrm{d}(L/L^*),
\label{Luminosityfunction}
\end{equation}
where $\alpha$ is the exponent at low luminosities, $\delta$ is the exponent at high luminosities, $\gamma$ determines the speed of the transition between the two power laws, and $L^*$ is the characteristic luminosity of the transition.  {The adopted parameters were $\alpha=-1.305$, $\delta=-7.13$, $\gamma=1.81$, and $M^{*}=-21.75$ (which corresponds to $L^{*}$), which is consistent with the luminosity function derived in \citet{Blanton:03}.}

After correcting for the luminosities, the luminosity-density field was then calculated on a cartesian grid using a 8\,$h^{-1}$Mpc B3-spline smoothing kernel,  {which corresponds to the supercluster scale \citep{Liivamägi2012}.} The density values ($D$) were normalized so that the densities are shown in units of the mean density of the field. Details of how the density field was calculated are given in \citet{Liivamägi2012}. Regions with the large-scale density $D>5.0$ can be defined as superclusters.

 {While calculating the density field, we also suppressed the finger-of-god redshift distortions using the rms sizes of galaxy groups on the sky and their rms radial velocities. \citet{Tempel:12} showed that this procedure makes the galaxy distribution in groups approximately spherical, as intended.}

 {We refer to \citet{Liivamägi2012} and \citet{Tempel:12} for more details of how the density field was calculated.}

\begin{figure}[t]
 \resizebox{\hsize}{!}{\includegraphics{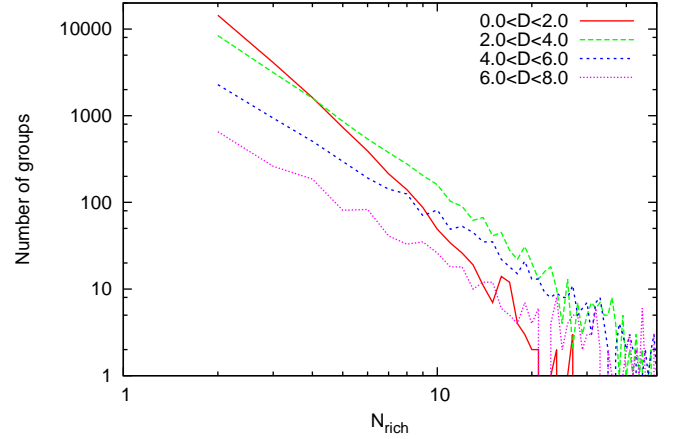}}
 \caption{The number of groups of different richnesses in four levels of large-scale density $D$. }
 \label{NGroups}
\end{figure}

Figure \ref{NGroups} shows the numbers of groups with different richnesses for different levels of $D$. Densities $0.0<D<2.0$ are characteristic of void regions, $2.0<D<4.0$ are typically filaments, $4.0<D<6.0$ are edges of superclusters, and the densities $6.0<D<8.0$ are supercluster core areas. The total number of galaxies in areas with $D>8.0$ is low, since the densest cores are small in volume. Because of this, we did not analyze these regions. Figure \ref{NGroups} shows that in void areas, groups are predominantly poorer than in denser large-scale environments.

\subsection{Spectral and morphological classification}

Active galaxies in the SDSS DR8 are divided into classes using their spectral emission lines and the criteria by \citet{Brinchmann2004}. The classification is based on emission-line ratios [\ion{O}{iii}]$/$H$\beta$ and [\ion{N}{ii}]$/$H$\alpha$. Galaxies with very weak or no emission lines are defined as unclassifiable. We used this class as our sample of passive galaxies. Of the galaxies with strong emission lines, star-forming galaxies and AGNs were distinguished by criteria found by \citet{Kewley2001} and \citet{Kauffmann2003}. In this work, we used the AGNs and star-forming galaxies, but not galaxies with composite spectra that have features between the two classes. We also omitted galaxies with classifications based on low signal-to-noise (S/N) data of either star-forming galaxy or AGN. Our data consists of 84\,427 passive galaxies, 89\,713 star-forming galaxies, and 9\,773 AGN.

We used the morphology classification of \citet{Tempel2011}, which takes into account the SDSS model fits, apparent ellipticities (and apparent sizes), and different galaxy colors. Additionally, the morphology is combined with the probabilities of being either elliptical or spiral galaxies given by \citet{Huertas2011}.  { \citet{Tempel:12}  showed that this classification agrees well with the \citet{Huertas2011} classification. In our classification, the galaxies were divided into spirals, ellipticals/S0, and galaxies with uncertain classifications.}  We rejected galaxies with uncertain classifications from our sample, and get 37\,683 passive ellipticals, 15\,614 passive spirals, 1\,513 star-forming ellipticals, and 78\,566 star-forming spirals.  {For about 47\% of the galaxies, the classification is unclear, the reason for which is twofold. Firstly, our classification in \citet{Tempel2011} was conservative and secondly, we used only galaxies, the \citet{Huertas2011} classification agreeing with our own. Hence, our classification is rather conservative and leaves out galaxies with uncertain data that may influence our analysis.}

According to \citet{Fukugita2004}, approximately 3\,\% of early-type galaxies are star-forming. On the other hand, \citet{Bamford2009} found that among the red (non-star-forming) galaxies, 67\,\% have an early-type morphology. Our data displays a similar trend: passive elliptical galaxies constitute approximately 70\,\% of the passive galaxies, and approximately 4\,\% of the elliptical galaxies are star-forming.

\begin{figure}
 \resizebox{\hsize}{!}{\includegraphics{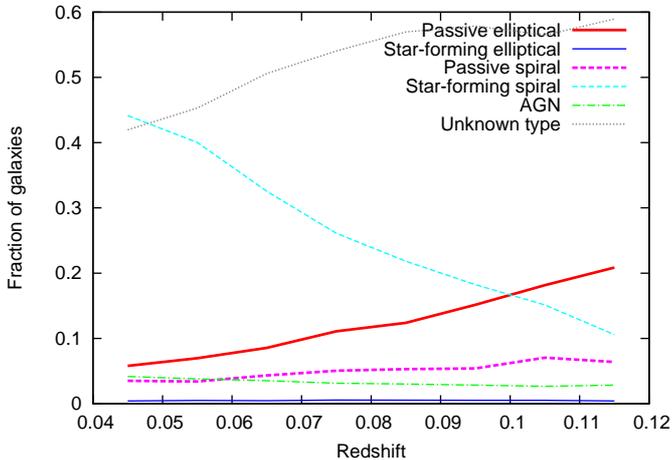}}
 \caption{Fractions of different types of galaxies as a function of  {redshift}.}
 \label{DistEffect}
\end{figure}

Figure \ref{DistEffect} shows the fractions of different types of galaxies at different  {redshifts}. The fractions were calculated in bins of $z=0.01$. The number of star-forming spiral galaxies declines strongly with increasing distance. In addition, the more distant galaxies are smaller and appear rounder in the sky, which makes it harder to classify the galaxies. Because of this, the number of galaxies with unknown morphology increases with distance. This again illustrates the aforementioned distance-dependent effects in our results caused by the magnitude-limited data. Since the passive elliptical galaxies tend to be more luminous than the star-forming spirals, they can be observed at larger distances. The distance range of our data, 120 to 340\,$h^{-1}$Mpc  {or 0.04 to 0.116 in redshift}, corresponds to $\sim0.6$ Gyr and is small enough to exclude there being any significant evolutionary effects within our analysis.

\section{Results}\label{Results}
\subsection{Large-scale environments}

The basic environmental properties of our five data samples are shown in Table \ref{samples}. This table gives the number of galaxies in each sample, the average colors of the galaxies, and the average group richness $N_{\mbox{rich}}$ and  {large-scale} density $D$ in the environment of the different types of galaxies. The data shown here includes galaxies with all possible group richnesses from field galaxies to clusters of more than a hundred galaxies.  {Errors are standard errors in the average.}
\begin{table*}
\centering
\begin{tabular}{c c c c c}
\hline\hline
Sample & N & $\langle g-r \rangle $ & $\langle N_{\mbox{rich}}\rangle$ & $\langle D\rangle$\\
\hline
Passive elliptical & 37683 & $0.796\pm0.001 $ & $10.0\pm0.2$ & $3.19\pm0.02 $ \\
Star-forming elliptical & 1513 & $0.70\pm 0.008$ & $4.2\pm0.3$ & $2.32\pm 0.05$\\
Passive spiral & 15614 & $0.776\pm0.002 $& $12.4\pm 0.3$ & $3.3\pm0.03$\\
Star-forming spiral & 78566 &$0.512\pm0.001$&$4.5\pm0.1$ & $2.1\pm0.01$ \\
AGN & 9773 & $0.737\pm0.002$ & $6.6\pm0.3$ & $2.58\pm 0.03$\\
\hline
\end{tabular}
\caption{Statistics of the galaxy samples: number of galaxies (N), average color ($\langle g-r \rangle $), average group richness ($\langle N_{\mbox{rich}}\rangle$), and average large-scale density ($\langle D\rangle$). }
\label{samples}
\end{table*}

Table \ref{samples} shows that passive galaxies with either elliptical or spiral morphologies occupy higher density environments than star-forming galaxies. Active galactic nuclei settle in-between: they have slightly richer environments than star-forming galaxies, but not as rich as the passive galaxies.  {\citet{Pasquali2009} also found that optically selected AGN reside in more massive halos than star-forming galaxies, but less massive ones than radio-emitting galaxies.} The differences can be seen on both, the group scale and the large scale. From these values alone, we are unable to identify the scale of the process that is the main cause of the differences.

\begin{figure}[t]
 \resizebox{\hsize}{!}{\includegraphics{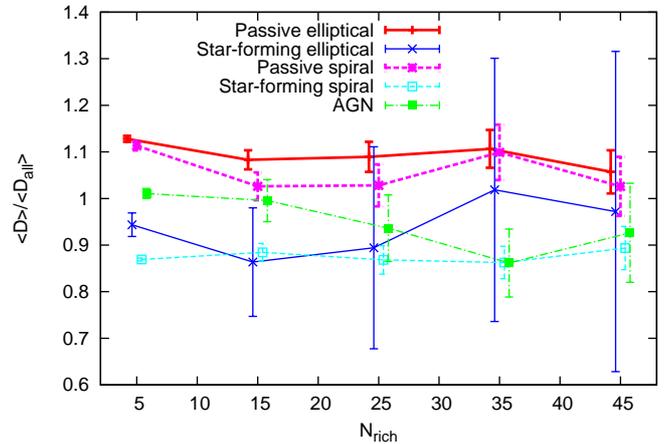}}
 \caption{ {Relative large-scale environments of different types of galaxies. The average large-scale density of each type divided by the average density of all galaxies is shown as a function of group richness. The error bars show Poisson errors}.}
 \label{RichVsDen}
\end{figure}

Figure \ref{RichVsDen} shows the average  relative large-scale density for the galaxy samples as a function of group richness. Galaxies are divided into bins of the richnesses of the groups where they reside with a binning of ten galaxies. The average  large-scale environmental density is calculated for galaxies  of each class in each bin.  {The density is divided by the average density of galaxies of all types in each richness bin. As a result, we see how the environments of different types differ from the average environment of all galaxies.}  With the same group richness, passive galaxies always have higher density environments than star-forming galaxies and AGNs. There is very little difference between the large-scale environments of spiral and elliptical passive galaxies.  {The similarity in the environments of passive galaxies of both morphological types is supported by the group-scale results of both the Galaxy Zoo project \citep{Bamford2009,Skibba2009} and the Space Telescope A901/2 Galaxy Evolution Survey (STAGES) \citep{Wolf2009}. }
For AGNs in the smallest groups, the  {large-scale} density is higher than for spiral, star-forming galaxies. In richer groups, AGNs and star-forming galaxies have approximately equal large-scale densities.

\begin{figure}[t]
 \resizebox{\hsize}{!}{\includegraphics{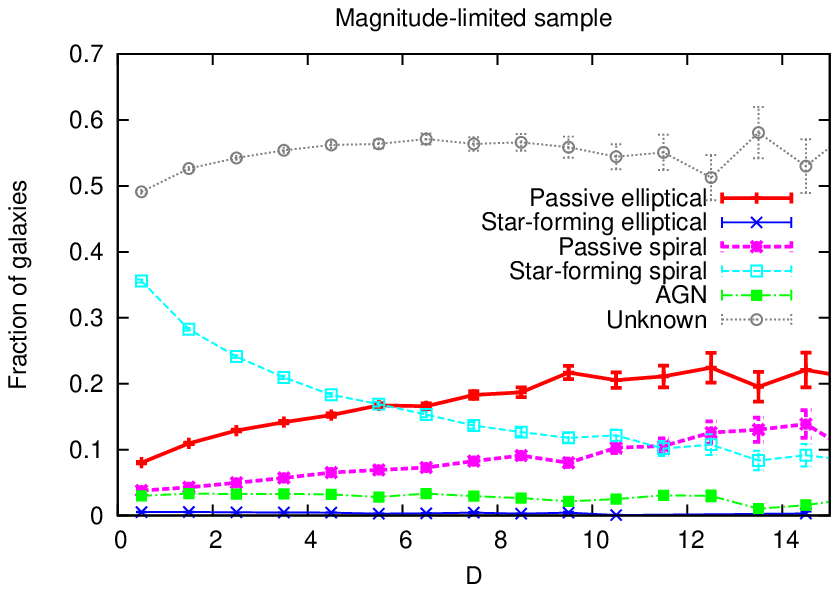}}\\
 \resizebox{\hsize}{!}{\includegraphics{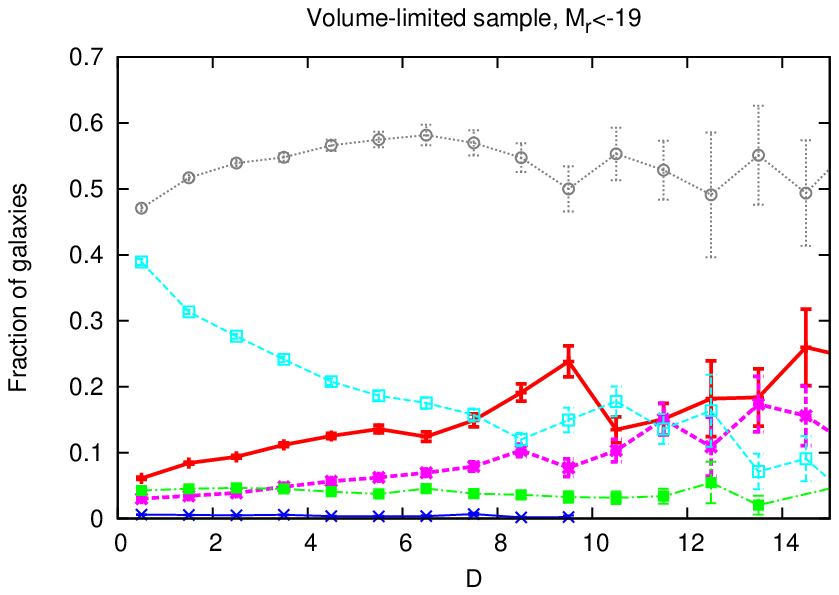}}
 \caption{Fractions of different samples of galaxies as a function of the  {large-scale} environmental density.  {Top panel shows the fractions for the complete magnitude-limited sample of galaxies between distances 120 and 340\,$h^{-1}$Mpc. Bottom panel shows the fractions in a volume-limited sample with M$_r<-19$ and distance between 120 and 225\,$h^{-1}$Mpc. } The numbers of galaxies are counted in bins of 1.0 density units.}
 \label{globFrac}
\end{figure}

The relative fractions of different types of galaxies in different  {large-scale} density levels are shown in Fig.~\ref{globFrac}. The number of galaxies of each type was counted in each bin of 1.0 mean densities. The fraction of each type was then calculated by dividing this number by the number of all galaxies in the bin (including galaxies with unknown classifications). All galaxies are more commonly in low-density regions, as was seen in Fig.~\ref{NGroups}. The error bars shown in Fig.~\ref{globFrac} are Poisson errors, and since the number of galaxies is high, the errors are small.  {The top panel of the figure shows the distribution of all galaxies in the magnitude-limited sample between distances of 120 and 340\,$h^{-1}$Mpc. To study the possible distance effects that may be caused by the magnitude-limited sample, we did the same calculation for a volume-limited sample (bottom panel of Fig.~\ref{globFrac}). The volume-limited sample was formed by a cut at absolute magnitude M$_r=-19$ and a distance cut at 225\,$h^{-1}$Mpc. There are no significant differences between the volume and magnitude-limited samples. }

The fractions of different types of galaxies presented in Fig.~\ref{globFrac} show that when moving towards higher  {large-scale} densities, the fraction of passive galaxies rises, while the fraction of star-forming galaxies drops. Passive elliptical galaxies become more numerous than spiral star-forming galaxies at the density level of five times the mean density, which is approximately the lower limit for supercluster regions. In addition, the fraction of passive spiral galaxies rises towards higher densities.

The fraction of AGN is equally small at all densities. According to the Kolmogorov-Smirnov test, the relation between the  {large-scale} density and the fraction of AGNs does not differ significantly from a random sample of galaxies of the same size. 

\begin{figure}[t]
 \resizebox{\hsize}{!}{\includegraphics{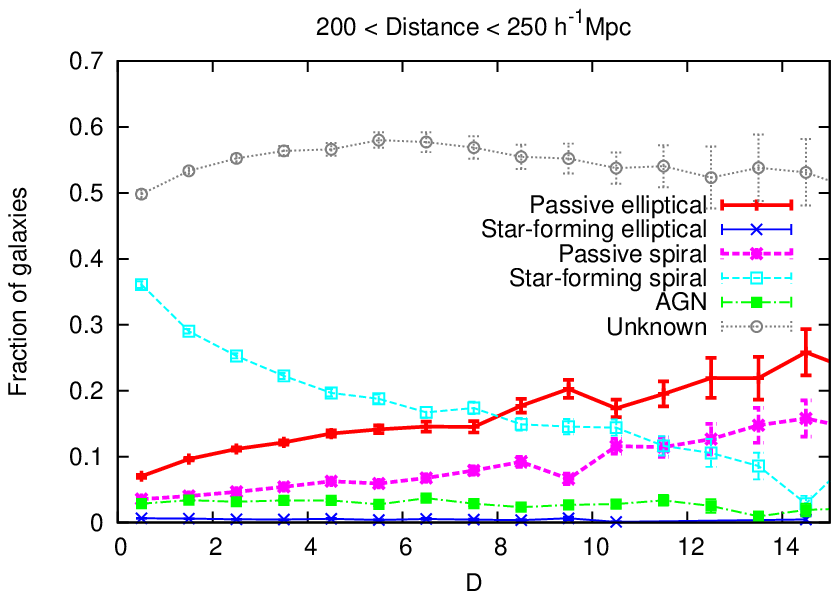}}\\
 \resizebox{\hsize}{!}{\includegraphics{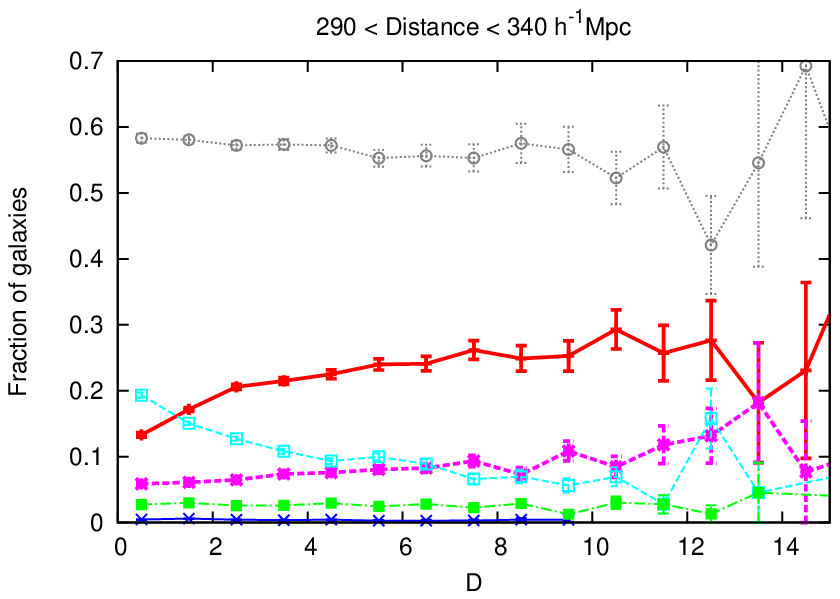}}
 \caption{Fractions of different samples of galaxies as a function of the  {large-scale} environmental density, as in Fig.~\ref{globFrac}, but for  {a magnitude-limited sample in} two distance bins of 200 to 250\,$h^{-1}$Mpc (top) and 290 to 340\,$h^{-1}$Mpc (bottom). }
 \label{globFracDist}
\end{figure}

Figure \ref{globFracDist} presents the effect of the distance-dependence of the fractions of different types of galaxies.  {The figure shows the fractions in the magnitude-limited sample as in Fig.~\ref{globFrac}, but separately for distances from 200 to 250\,$h^{-1}$Mpc, and from 290 to 340\,$h^{-1}$Mpc.} The results are qualitatively similar: the fraction of star-forming galaxies declines and the fraction of passive galaxies rises when moving towards higher densities.  This implies that although the exact values of the fractions of different types of galaxies depend on the sample volume, the phenomenon of observed environmental dependence is not caused by distance-related selection effects.  {In the larger distance interval, 290 to 340\,$h^{-1}$Mpc, there are fewer rich superclusters \citep{Einasto2012a}, and so the number of galaxies in the highest densities is smaller and random fluctuations larger.}

\subsection{Group properties and environment}

\begin{figure}
 \resizebox{\hsize}{!}{\includegraphics{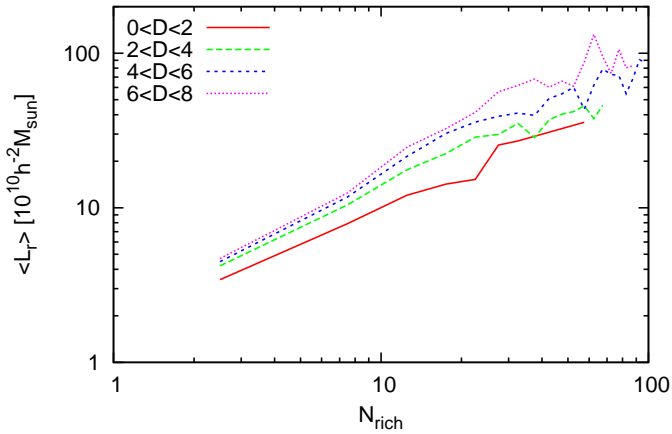}}
 \caption{Average luminosity for groups as a function of group richness in four large-scale density bins.}
 \label{NvsL}
\end{figure}

To study the group-scale effects on galaxies, we used the richness $N_{\mbox{rich}}$ and  {total} luminosity $L_r$ of the group as measures. They are correlated, but we can see in Fig.~\ref{NvsL} that their relation depends on the large-scale density. This figure shows the average luminosity of groups  {as a function of group richness} in four large-scale density bins.  {Luminosities were calculated in richness bins of five galaxies.} {Groups in environments of higher large-scale density are on average more luminous than groups in low-density environments when their richnesses are the same.}  {This seems to imply that halo mass, which is strongly correlated with richness, does not completely explain the total luminosities of groups. The mass-to-luminosity ratio may depend on the large-scale environment. The main reason for this result is that the most luminous galaxies are more often in high-density environments. We showed in Fig.~\ref{RichVsDen} that passive galaxies are on average in higher-density large-scale environments than star-forming spirals. Since the most luminous galaxies are passive ellipticals, it is expected that the average luminosities are higher in high-density regions. }

\begin{figure*}[t]
 \centering
 \begin{tabular}{c c}
 \resizebox{\columnwidth}{!}{\includegraphics{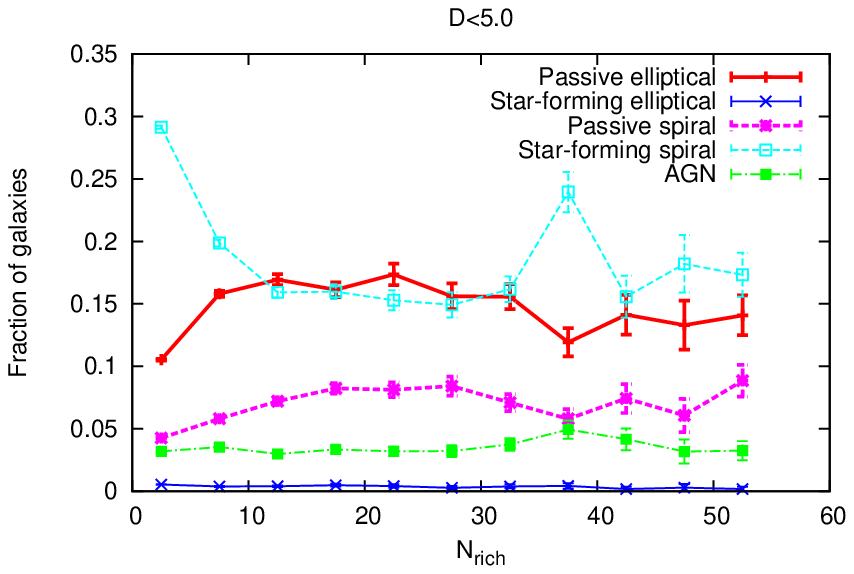}}&
 \resizebox{\columnwidth}{!}{\includegraphics{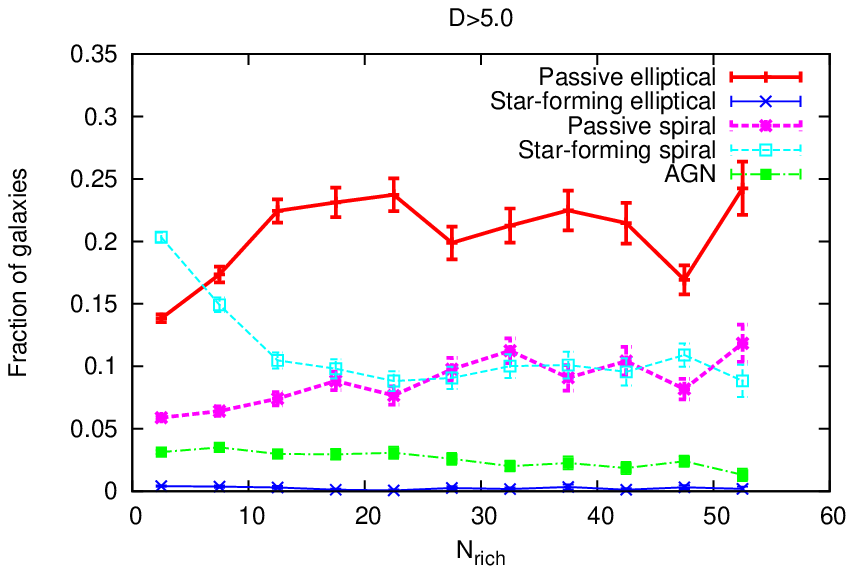}}
 \end{tabular}
 \caption{ {Fractions of different samples of galaxies as a function of group richness in regions with lower-density large-scale environments (left) and in superclusters (right). Error bars show the Poisson errors. }}
 \label{fracHighLow}
\end{figure*}

 {Figure \ref{fracHighLow} shows the fractions of different samples of galaxies as a function of the group richness. The fractions are calculated in richness bins of five galaxies. The left-hand plot shows the results for galaxies in  {large-scale} environments of density $D<5.0$, while the right-hand plots show the results for galaxies in supercluster regions of density $D>5.0$.} 

%In low-density environment, the number of galaxies decreases rapidly when moving towards richer groups. Most galaxies in voids are preferentially in very small groups. In supercluster environments, there are relatively more galaxies in richer groups.

The fractions of different types of galaxies depend on their group-scale environment in the same way as their large-scale environment: the richer the group, the more passive galaxies it contains.  {This rise in the fraction of passive (red) galaxies as a function of richness in small groups was also found by \citet{Hansen2009} and \citet{Weinmann2006}.} However, there is a difference between the supercluster environments and the low-density environments that can be seen even when the local environment is the same. In low-density environments, the numbers of passive elliptical and star-forming spiral galaxies are equal in groups with ten or more galaxies, while in the supercluster environments the fraction of elliptical galaxies becomes larger.  {This result is in accordance with the results by \citet{Tempel2011} and \citet{Einasto2007}}. The fraction of passive spiral galaxies rises moderately with the group richness, but it does not depend on the large-scale environment. The fraction of AGNs slightly decreases in the supercluster environment, while in void or filament environments it is constant. This was also seen in Fig.~\ref{RichVsDen}.

\begin{figure}[t]
 \centering
 \resizebox{\hsize}{!}{\includegraphics{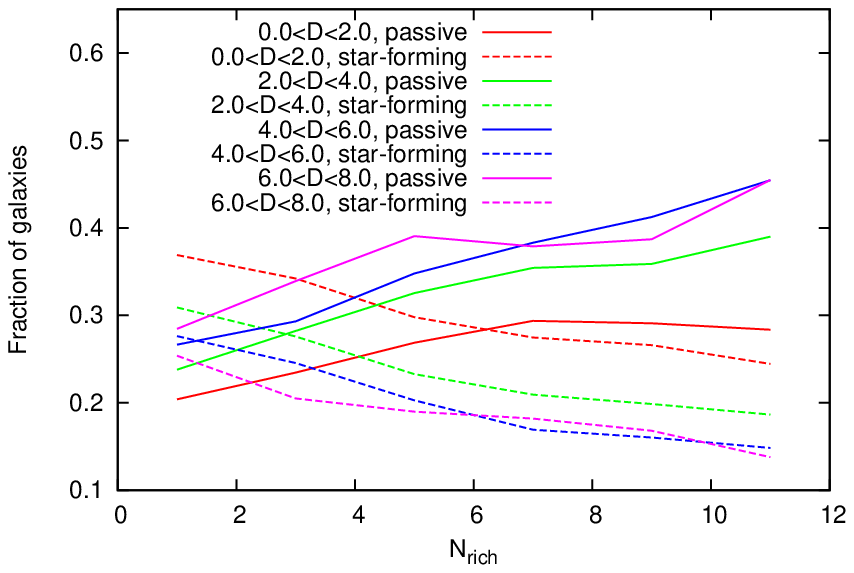}}\\
 \resizebox{\hsize}{!}{\includegraphics{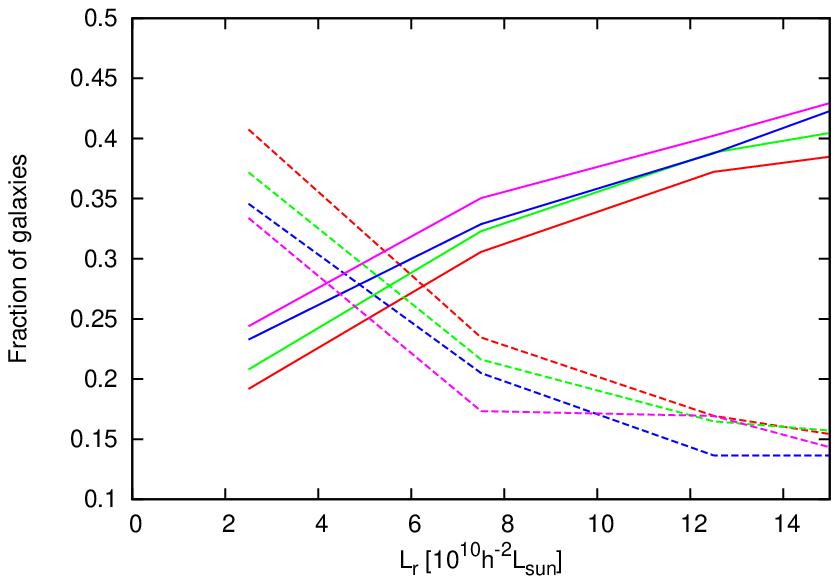}}
 \caption{Fractions of passive (solid lines) and star-forming (dashed line) galaxies as functions of group richness (top) and luminosity (bottom) in four large-scale density bins. The fractions are calculated in richness bins of two galaxies  {and luminosity bins of $5\times 10^{10}h^{-2}L_\odot$}.}
 \label{FracSfPass}
\end{figure}

The most prominent dependence on the environment is the difference between star-forming and passive galaxies. To study this in more detail, we calculated the fractions of all star-forming and passive galaxies regardless of their morphology. We divided these galaxies into four bins based on their large-scale environment: $0.0<D<2.0$ (typically voids), $2.0<D<4.0$ (filaments), $4.0<D<6.0$ (edges of superclusters), and $6.0<D<8.0$ (supercluster cores). The fractions of star-forming and passive galaxies as functions of group richness and group luminosity in these density bins are shown in Fig.~\ref{FracSfPass}. To smooth the curves and improve the reliability, we used a binning of two galaxies in group richness. In all the bins of all samples up to a group richness of ten galaxies, the numbers of galaxies exceed a hundred.  For luminosities, the fractions were calculated in bins of $5\times 10^{10}h^{-2}L_\odot$. 

The crossing point at which passive galaxies become more numerous than star-forming galaxies moves to lower group richness as the  {large-scale} density grows. In low-density regions in groups of more than eight galaxies the fractions of passive and star-forming galaxies are approximately equal, while in higher density regions the fraction of passive galaxies gets higher, as the fraction of star-forming galaxies gets lower. When the  {large-scale} density gets close to the supercluster limit ($D\sim5$), increasing the density does not change the fractions any further. The dependence of the crossing point on the large-scale density can also been seen in terms of luminosity, although the differences between the density levels are smaller.

 {The differences between the supercluster and low-density environments may be due to the different evolutions of galaxies in high and low-density environments.} The properties of galaxies that are characteristic of galaxies in the early phases of their evolution are more often found in low-density environments. In Figs. \ref{fracHighLow} and \ref{FracSfPass}, we can see that the group scale affects galaxies in the smallest groups. In richer groups, the fractions of different types of galaxies are almost independent of group richness. A possible explanation of this is that in the small groups, the evolution is ongoing. Larger groups have been formed through the mergers of smaller groups, which are older and more mature structures.

\begin{figure}[t]
 \resizebox{\hsize}{!}{\includegraphics{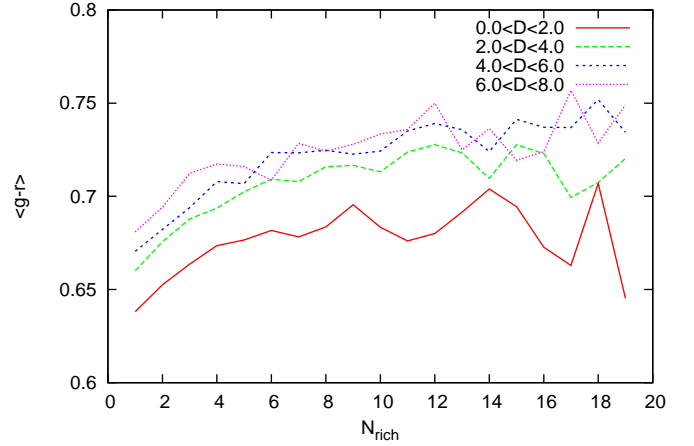}}
 \caption{Average $g-r$ color of galaxies as a function of richness of their group for galaxies in four bins of large-scale density.}
 \label{color}
\end{figure}

We also calculated average $g-r$ colors of galaxies in different large-scale environments as a function of group richness. The results are shown in Fig.~\ref{color} for all galaxies, divided into the four bins of large-scale density. On average, galaxies are approximately 0.05 magnitudes redder in supercluster areas than in voids.  The transition from star-forming to passive galaxies can be seen in terms of galaxy colors. The average color gets redder when the group richness rises to $\sim$10 galaxies. For richer groups, the color is independent of group richness, indicating that the formation of these groups is more or less finished. However, this does not mean that they are virialised, but that they can still contain substructures \citep{Einasto2012}.
The color-dependency found in Fig.~\ref{color} is caused by changes in the relative fractions of star-forming and passive galaxies. The colors of different types of galaxies do not depend on their environment on either the group or supercluster scale. The average colors are those presented in Table \ref{samples} regardless of environment.  {Therefore, the dependence of color on environment is another way of expressing the dependence of star-forming activity on environment. Our results are consistent with other studies based on colors of galaxies \citep{Hansen2009,Skibba2009b}. }

\subsection{Distribution of galaxies in groups}

 {To study the possible differences in the radial distributions of elliptical
and spiral galaxies in groups, we analyzed their
three-dimensional spatial distribution
around the center of the group. We used the given group
  distance and the projected coordinates to calculate the group center ($x_c$,
  $y_c$, $z_c$), after which we measured the separation from this
  point for all the galaxies in the group. For galaxy distances, we
  used the co-moving distances where the finger-of-god effect is suppressed.
To study all groups together and evaluate the possible differences
between galaxy types, we calculated the mean value of the separations
for all galaxies in the groups.
We first selected the data based on the surrounding large-scale density
 {$D<5.0$}, and divided the sample into two categories of distance:
groups that are closer than 200\,$h^{-1}$Mpc and groups that are
beyond this distance.
Our analysis shows that there is no difference between the mean
distances of galaxies in the groups as a function of group distance.
This is a good sign as this shows that there is no bias in the galaxy
type distribution that would depend on the distance.
All values we refer to in the text later on, are for the whole
  sample. If we further divide the group sample into passive and star-forming
galaxies and add another environment $D \ge 5.0$, we have
four different galaxy populations in two different large-scale
environments.
The results show that elliptical galaxies and spiral galaxies are
distributed in different ways.
In the low density sample, passive elliptical galaxies have a mean
separation of $0.307\pm0.001$\,$h^{-1}$Mpc, while the passive spiral
galaxies are
further away with the mean distance of $0.344\pm0.003$\,$h^{-1}$Mpc.
In the high density region the mean separations are larger:
$0.617\pm0.006$\,$h^{-1}$Mpc for elliptical galaxies and
$0.705\pm0.01$\,$h^{-1}$Mpc for spiral galaxies.
The same feature is also observed among star-forming galaxies and the
referred values are approximately the same.}

 {To study whether the mean separations depend on the number
  of members, we performed another test.
The same surrounding densities were used, and groups of galaxies were
divided into a poor sample with $N_{\mbox{rich}}=3$ or 5 and rich sample
with $N_{\mbox{rich}}\ge 10$.
The result was that in general both elliptical galaxies and spiral
galaxies have larger mean distances in rich groups.
For $N_{\mbox{rich}}=3$, the mean distance is $\sim 0.2\,h^{-1}$Mpc and
for $N_{\mbox{rich}}=5$ the distance is $\sim 0.3\,h^{-1}$Mpc and for
$N_{\mbox{rich}} \ge 10$ the distance is already  $\sim
0.5-1.0\,h^{-1}$Mpc.
This was expected since richer groups are typically larger in size. In these samples, the mean distances are also larger for spiral
galaxies than elliptical galaxies.
If we fix the number of members in the group, the separations do not depend as much on the large-scale
density as before. {Only in the $N_{\mbox{rich}}\ge 10$ sample are there notable differences in the mean distances of galaxies at different large-scale densities. In high-density environments, there are more rich groups than in the low-density environments, and therefore the mean distances of galaxies are larger in high- than in low-density environments.}
These are interesting results showing the importance of group richness in determining the mean distances of elliptical and spiral galaxies.
Passive and starforming galaxies behave practically in the same way
in most samples. Again, only for rich groups are there systematic differences in the mean distances. Star-forming galaxies have $\sim$16\,\% larger distances than those for passive galaxies. }

\section{Discussion}\label{Discussion}

Our results support the conclusions of \citet{Bamford2009} and \citet{vanderWel2010} that color and  {structural parameters} of galaxies depend differently on the environment. Star-forming galaxies occupy groups of lower richness and  {large-scale} density than passive galaxies regardless of their morphology. For spiral galaxies, the dependence of star-formation rate on the environment was also observed by \citet{Koopmann2004}. On the basis of these studies, we can conclude that what we observe as the morphology-density relation is mostly due to spiral galaxies being usually star-forming and elliptical galaxies passive. 

Several mechanisms have been suggested for quenching star formation in galaxies. 
These include interactions between the galaxy and the surrounding 
intracluster medium \citep[see the review by][]{vanGorkom2004} and mergers of 
galaxies \citep{Hopkins2008}. Mergers are expected to happen 
in small groups. \citet{SolAlonso2006} found that galaxy interactions 
are effective at triggering star-formation activity in low- and moderate-density environments.
The different distributions of star-forming and passive galaxies in low and high-density environments implies that environments can affect the quenching of star formation. We found that passive galaxies with both elliptical and spiral morphologies prefer richer groups and higher large-scale densities than star-forming spirals. This result indicates that the environmental effects can quench star formation both by changing the morphology of the galaxy from spiral to elliptical, and without affecting the morphology.
 {\citet{Kauffmann2004} concluded that since the morphology does not depend on the environment, the trend in star formation cannot be driven by processes that alter structure, such as mergers or harassment.}

According to our results, the fraction of AGNs in all galaxies does not strongly depend on either group richness or large-scale environment. In \citet{Lietzen2011}, radio quiet quasars and Seyfert galaxies, which were selected by the same criteria as we used here, were distributed predominantly in low-density environments. Based on the results in this paper, radio-quiet AGNs may reflect the overall distribution of galaxies. \citet{Lietzen2009} found that quasars have fewer neighboring galaxies than luminous, inactive galaxies. The difference between their result and the one in this paper is caused by their control sample, which consisted of very luminous galaxies. The most luminous galaxies  are more likely to be located in high-density environments. On the other hand, our results agree with those of \citet{Silverman2009}, who found that the fraction of group galaxies hosting an AGN is similar to the fraction of field galaxies, and \citet{Martini2007}, who found that AGN do not appear to have different substructure distributions from inactive galaxies in a cluster. 
The flat distribution of AGNs ( {in Figs. \ref{globFrac} and \ref{fracHighLow}}) suggests that the 
environment does not have a strong effect on triggering AGNs. We found small 
differences between the average large-scale densities as a function of the group 
richness of AGNs and other types of galaxies
(Fig.~\ref{RichVsDen}). 
 {This result may indicate that in regions with high large-scale density, AGNs would be more likely in smaller groups than in regions of low large-scale density}.

Several studies indicate that galaxy properties also depend on the large-scale environment in which they reside \citep[e.g.][]{Park2008, Cooper2010}. In these studies, the large-scale environment is often interpreted as relatively small regions around the object 
under study, from kpc to a few Mpc.  {Here the large-scale environment refers to different density levels of the large-scale galaxy density field and its characteristics, superclusters and voids ranging scales from ten to a hundred Mpc. We found that the fractions of different types of galaxies depend on both group richness and the large-scale density. \emph{Galaxies in equally rich groups are more likely to be passive if they are in supercluster environments than if they are in low density region (Figs. \ref{fracHighLow} and \ref{FracSfPass})}. This result implies that the large-scale environment affects both the galaxy and group properties.}
% {Moreover we may For small groups (less than ten galaxies) group 
%richness is an important factor, while in richer groups fraction 
%of different types of galaxies 
%does not longer depends on local richness so much}.
%The  {large-scale} environmental density affects the fractions of different 
%galaxies independently of the group richness. 
A similar result was found by \citet{Einasto2008} for individual rich superclusters. 
They reported that in equally rich groups there are more early-type galaxies in supercluster cores than in the outskirts of superclusters. In addition, the galaxies that do not belong to any group are more likely to be early-type if they are in supercluster cores. 

 {Using spectrocopically identified compact-group galaxies, \citet{Scudder2012} showed that star-formation rates of star-forming galaxies differ significantly between isolated groups and groups embedded within  larger systems. {Galaxies in isolated groups have higher star-formation rates than those  in groups of the same stellar mass in larger systems.} On the basis of a statistical study of galaxies in voids and void walls in the SDSS and 2dFGRS catalogs, \citet{Ceccarelli2008} report that the galaxy population is also affected by the large-scale modulation of star formation in galaxies. \citet{Sorrentino2006} studied galaxy properties in different large-scale environments in the SDSS DR4. They concluded that the fraction of late-type galaxies decreases and the fraction of early-type galaxies increases from underdense to denser environments. This transition reflects a smooth continuity of galaxy properties from voids to clusters. This implies that the method through which the environment affects the galaxies changes gradually with increasing density rather than being a phenomenon tied to a certain threshold density.}

\citet{Einasto2005a} showed by numerical simulations that in void
regions only poor systems are formed, and that these systems grow very
slowly. Most of the particles are maintained in their primordial form,
remaining non-clustered. In superclusters, the dynamical evolution is
rapid and starts earlier, consisting of a continua of transitions of dark
matter particles to building blocks of galaxies, groups, and
clusters. Moreover, dark matter halos in voids are found to be less
massive, less luminous, and their growth of mass is suppressed and
stops at earlier epoch than in high-density regions. The merger rate
is a function of the environmental density \citep{Gottlöber2003,
Colberg2005, Einasto2005a, Fakhouri2009}. \citet{Tempel2009} showed 
that during the structure formation of the Universe, 
halo sizes in the supercluster core regions increase by 
many factors, while in void regions, halo sizes remain 
unchanged. Thus, during the evolution of structure 
the overall density in voids decreases, which supresses the evolution of the small-scale protohalos in voids
and low density regions.

 {\citet{Keres2005} used hydrodynamical simulations to study the hot and cold gas supplies to the galaxies. They concluded that cold and hot gas modes depend on galaxy redshift and environment. The cold gas that was prevalent at high redshifts and in low density regions nowadays may make important contributions to the star-formation rates of low mass systems. \citet{Cowie1996} found evidence of cosmic downsizing: during the structure formation galaxies  transform from being star-forming to passive. \citet{Cen2011} showed through hydrodynamical simulations that for galaxies at $z=0$ star formation is efficient in low density regions while it is substantially suppressed in cluster environments. {During hierarchical structure formation, gas is heated in high density regions (groups, clusters, and superclusters), and a continuously larger fraction of the gas has entered too hot a phase to feed the residing galaxies.} As a result, cold gas supply to galaxies in these regions is suppressed. The net effect is that star formation gradually shifts from the larger halos that populate overdense regions to  lower density environments. Thus, a lack of cold gas due to gravitational heating in dense regions may provide a physical explanation of cosmic downsizing and one of its manifestations is observed as the color-density relation.}

 {The evolution predicted by numerical simulations might be consistent with the observational results of this paper. 
%If the evolution of galaxies in the underdense regions is suppressed (i.e. their evolution is slower and they are younger), relatively large group is required to turn galaxies from star-forming to passive stage. Available amount and mixture of the hot and cold gas which seems to depend on galaxy mass and external environment are in undoubtedly in vital role here. 
Our results imply that the environment may affect the galaxy evolution and properties on two overlapping levels. On the first level, the properties are affected in the large-scale density field, which affects in turn the evolution of both galaxy and group sized halos  depending on the field density. At the second, more local level, the properties depend on the gas supply together with baryonic and gravitational processes acting in groups that host galaxies. Figure \ref{FracSfPass} can also be interpreted as evidence that galaxies in lower-density regions developed later than galaxies in similar mass groups in high-density environments. This can be studied in more detail through the assembly history of groups in numerical simulations.}

\section{Summary and conclusions}\label{Conclusions}

We have studied the environments of galaxies of the SDSS DR8 at distances of between 120 and 340\,$h^{-1}$Mpc. We divided the galaxies into different samples based on their spectral properties and morphology. We measured the large-scale environment using a luminosity-density field and the group-scale environment with the group richness. Our main results are the following:
\begin{itemize}
\item{Passive galaxies are located in denser large-scale environments than star-forming galaxies. There is no significant difference between passive elliptical and passive spiral galaxies. }
\item{The fraction of galaxies that are star-forming declines and the fraction of passive galaxies rises as the richness of a group rises from one to approximately ten galaxies. In groups with richnesses of between 20 and 50 galaxies, the fractions of galaxies of different types do not depend on group richness. }
\item{The group richness at which passive galaxies become more numerous than  star-forming galaxies depends on the large-scale  environment. When the large-scale density grows from levels typical  of voids to supercluster regions, the group richness where most galaxies become passive  declines. In addition, the fractions of star-forming and passive galaxies in rich groups   depend on the large-scale environment: {in voids, the fractions of  passive and star-forming galaxies are approximately equal to each other in groups with more than ten  galaxies, while in superclusters the fractions of passive galaxies are considerably larger than those of star-forming galaxies.}}
\item{Equally rich groups are more luminous in supercluster regions than in voids.}
\item{The fraction of galaxies with an AGN does not depend strongly on either group richness or the large-scale density.}

\end{itemize}

We conclude that galaxy evolution is affected by both the group where the galaxy resides and its large-scale environment. In the future we plan to use numerical simulations to study the plausible baryonic and gravitational processes that determine galaxy evolution in groups of galaxies in different large-scale environments. Another invaluable approach will be to study in more detail the properties that are known to characterise groups e.g. their \ion{H}{i} content, stellar mass, and X-ray properties.

\begin{acknowledgements}

We thank the referee, Ramin Skibba, for his useful comments that greatly helped to improve this paper.

H. Lietzen was supported by the Finnish Cultural Foundation and Emil Aaltonen's foundation.

The present study was supported by the Estonian Science Foundation
grants No. 8005, 7765, 9428, and MJD272, by the Estonian Ministry for Education and Science research project SF0060067s08, and by the European Structural Funds grant for the Centre of Excellence "Dark Matter in (Astro)particle Physics and Cosmology" TK120.

    Funding for the SDSS and SDSS-II has been provided by the Alfred P. Sloan Foundation, the Participating Institutions, the National Science Foundation, the U.S. Department of Energy, the National Aeronautics and Space Administration, the Japanese Monbukagakusho, the Max Planck Society, and the Higher Education Funding Council for England. The SDSS Web Site is http://www.sdss.org/.

    The SDSS is managed by the Astrophysical Research Consortium for the Participating Institutions. The Participating Institutions are the American Museum of Natural History, Astrophysical Institute Potsdam, University of Basel, University of Cambridge, Case Western Reserve University, University of Chicago, Drexel University, Fermilab, the Institute for Advanced Study, the Japan Participation Group, Johns Hopkins University, the Joint Institute for Nuclear Astrophysics, the Kavli Institute for Particle Astrophysics and Cosmology, the Korean Scientist Group, the Chinese Academy of Sciences (LAMOST), Los Alamos National Laboratory, the Max-Planck-Institute for Astronomy (MPIA), the Max-Planck-Institute for Astrophysics (MPA), New Mexico State University, Ohio State University, University of Pittsburgh, University of Portsmouth, Princeton University, the United States Naval Observatory, and the University of Washington.

\end{acknowledgements}

\bibliographystyle{aa}
\bibliography{references.bib}

\begin{thebibliography}{73}
\expandafter\ifx\csname natexlab\endcsname\relax\def\natexlab#1{#1}\fi

\bibitem[{{Aihara} {et~al.}(2011){Aihara}, {Allende Prieto}, {An}, {Anderson},
  {Aubourg}, {Balbinot}, {Beers}, {Berlind}, {Bickerton}, {Bizyaev}, {Blanton},
  {Bochanski}, {Bolton}, {Bovy}, {Brandt}, {Brinkmann}, {Brown}, {Brownstein},
  {Busca}, {Campbell}, {Carr}, {Chen}, {Chiappini}, {Comparat}, {Connolly},
  {Cortes}, {Croft}, {Cuesta}, {da Costa}, {Davenport}, {Dawson}, {Dhital},
  {Ealet}, {Ebelke}, {Edmondson}, {Eisenstein}, {Escoffier}, {Esposito},
  {Evans}, {Fan}, {Femen{\'{\i}}a Castell{\'a}}, {Font-Ribera}, {Frinchaboy},
  {Ge}, {Gillespie}, {Gilmore}, {Gonz{\'a}lez Hern{\'a}ndez}, {Gott}, {Gould},
  {Grebel}, {Gunn}, {Hamilton}, {Harding}, {Harris}, \& {et al.}}]{Aihara:11}
{Aihara}, H., {Allende Prieto}, C., {An}, D., {et~al.} 2011, \apjs, 193, 29

\bibitem[{{Baldry} {et~al.}(2006){Baldry}, {Balogh}, {Bower}, {Glazebrook},
  {Nichol}, {Bamford}, \& {Budavari}}]{Baldry2006}
{Baldry}, I.~K., {Balogh}, M.~L., {Bower}, R.~G., {et~al.} 2006, \mnras, 373,
  469

\bibitem[{{Balogh} {et~al.}(2004){Balogh}, {Baldry}, {Nichol}, {Miller},
  {Bower}, \& {Glazebrook}}]{Balogh2004}
{Balogh}, M.~L., {Baldry}, I.~K., {Nichol}, R., {et~al.} 2004, \apjl, 615, L101

\bibitem[{{Bamford} {et~al.}(2009){Bamford}, {Nichol}, {Baldry}, {Land},
  {Lintott}, {Schawinski}, {Slosar}, {Szalay}, {Thomas}, {Torki}, {Andreescu},
  {Edmondson}, {Miller}, {Murray}, {Raddick}, \& {Vandenberg}}]{Bamford2009}
{Bamford}, S.~P., {Nichol}, R.~C., {Baldry}, I.~K., {et~al.} 2009, \mnras, 393,
  1324

\bibitem[{{Berlind} {et~al.}(2006){Berlind}, {Frieman}, {Weinberg}, {Blanton},
  {Warren}, {Abazajian}, {Scranton}, {Hogg}, {Scoccimarro}, {Bahcall},
  {Brinkmann}, {Gott}, {Kleinman}, {Krzesinski}, {Lee}, {Miller}, {Nitta},
  {Schneider}, {Tucker}, \& {Zehavi}}]{Berlind:06}
{Berlind}, A.~A., {Frieman}, J., {Weinberg}, D.~H., {et~al.} 2006, \apjs, 167,
  1

\bibitem[{{Blanton} \& {Berlind}(2007)}]{Blanton2007}
{Blanton}, M.~R. \& {Berlind}, A.~A. 2007, \apj, 664, 791

\bibitem[{{Blanton} {et~al.}(2005){Blanton}, {Eisenstein}, {Hogg}, {Schlegel},
  \& {Brinkmann}}]{Blanton2005}
{Blanton}, M.~R., {Eisenstein}, D., {Hogg}, D.~W., {Schlegel}, D.~J., \&
  {Brinkmann}, J. 2005, \apj, 629, 143

\bibitem[{{Blanton} {et~al.}(2003){Blanton}, {Hogg}, {Bahcall}, {Brinkmann},
  {Britton}, {Connolly}, {Csabai}, {Fukugita}, {Loveday}, {Meiksin}, {Munn},
  {Nichol}, {Okamura}, {Quinn}, {Schneider}, {Shimasaku}, {Strauss}, {Tegmark},
  {Vogeley}, \& {Weinberg}}]{Blanton:03}
{Blanton}, M.~R., {Hogg}, D.~W., {Bahcall}, N.~A., {et~al.} 2003, \apj, 592,
  819

\bibitem[{{Blanton} \& {Roweis}(2007)}]{Blanton:07}
{Blanton}, M.~R. \& {Roweis}, S. 2007, \aj, 133, 734

\bibitem[{{Brinchmann} {et~al.}(2004){Brinchmann}, {Charlot}, {White},
  {Tremonti}, {Kauffmann}, {Heckman}, \& {Brinkmann}}]{Brinchmann2004}
{Brinchmann}, J., {Charlot}, S., {White}, S.~D.~M., {et~al.} 2004, \mnras, 351,
  1151

\bibitem[{{Ceccarelli} {et~al.}(2008){Ceccarelli}, {Padilla}, \&
  {Lambas}}]{Ceccarelli2008}
{Ceccarelli}, L., {Padilla}, N., \& {Lambas}, D.~G. 2008, \mnras, 390, L9

\bibitem[{{Cen}(2011)}]{Cen2011}
{Cen}, R. 2011, \apj, 741, 99

\bibitem[{{Colberg} {et~al.}(2005){Colberg}, {Sheth}, {Diaferio}, {Gao}, \&
  {Yoshida}}]{Colberg2005}
{Colberg}, J.~M., {Sheth}, R.~K., {Diaferio}, A., {Gao}, L., \& {Yoshida}, N.
  2005, \mnras, 360, 216

\bibitem[{{Cooper} {et~al.}(2010){Cooper}, {Gallazzi}, {Newman}, \&
  {Yan}}]{Cooper2010}
{Cooper}, M.~C., {Gallazzi}, A., {Newman}, J.~A., \& {Yan}, R. 2010, \mnras,
  402, 1942

\bibitem[{{Cowie} {et~al.}(1996){Cowie}, {Songaila}, {Hu}, \&
  {Cohen}}]{Cowie1996}
{Cowie}, L.~L., {Songaila}, A., {Hu}, E.~M., \& {Cohen}, J.~G. 1996, \aj, 112,
  839

\bibitem[{{Dressler}(1980)}]{Dressler1980}
{Dressler}, A. 1980, \apj, 236, 351

\bibitem[{{Einasto} {et~al.}(2003){Einasto}, {H{\"u}tsi}, {Einasto}, {Saar},
  {Tucker}, {M{\"u}ller}, {Hein{\"a}m{\"a}ki}, \& {Allam}}]{Einasto2003}
{Einasto}, J., {H{\"u}tsi}, G., {Einasto}, M., {et~al.} 2003, \aap, 405, 425

\bibitem[{{Einasto} {et~al.}(2005{\natexlab{a}}){Einasto}, {Tago}, {Einasto},
  {Saar}, {Suhhonenko}, {Hein{\"a}m{\"a}ki}, {H{\"u}tsi}, \&
  {Tucker}}]{Einasto2005a}
{Einasto}, J., {Tago}, E., {Einasto}, M., {et~al.} 2005{\natexlab{a}}, \aap,
  439, 45

\bibitem[{{Einasto} \& {Einasto}(1987)}]{Einasto1987}
{Einasto}, M. \& {Einasto}, J. 1987, \mnras, 226, 543

\bibitem[{{Einasto} {et~al.}(2007){Einasto}, {Einasto}, {Tago}, {Saar},
  {Liivam{\"a}gi}, {J{\~o}eveer}, {H{\"u}tsi}, {Hein{\"a}m{\"a}ki},
  {M{\"u}ller}, \& {Tucker}}]{Einasto2007}
{Einasto}, M., {Einasto}, J., {Tago}, E., {et~al.} 2007, \aap, 464, 815

\bibitem[{{Einasto} {et~al.}(2012{\natexlab{a}}){Einasto}, {Liivam{\"a}gi},
  {Tempel}, {Saar}, {Vennik}, {Nurmi}, {Gramann}, {Einasto}, {Tago},
  {Hein{\"a}m{\"a}ki}, {Ahvensalmi}, \& {Mart{\'{\i}}nez}}]{Einasto2012a}
{Einasto}, M., {Liivam{\"a}gi}, L.~J., {Tempel}, E., {et~al.}
  2012{\natexlab{a}}, \aap, 542, A36

\bibitem[{{Einasto} {et~al.}(2008){Einasto}, {Saar}, {Mart{\'{\i}}nez},
  {Einasto}, {Liivam{\"a}gi}, {Tago}, {Starck}, {M{\"u}ller},
  {Hein{\"a}m{\"a}ki}, {Nurmi}, {Paredes}, {Gramann}, \&
  {H{\"u}tsi}}]{Einasto2008}
{Einasto}, M., {Saar}, E., {Mart{\'{\i}}nez}, V.~J., {et~al.} 2008, \apj, 685,
  83

\bibitem[{{Einasto} {et~al.}(2005{\natexlab{b}}){Einasto}, {Suhhonenko},
  {Hein{\"a}m{\"a}ki}, {Einasto}, \& {Saar}}]{Einasto2005}
{Einasto}, M., {Suhhonenko}, I., {Hein{\"a}m{\"a}ki}, P., {Einasto}, J., \&
  {Saar}, E. 2005{\natexlab{b}}, \aap, 436, 17

\bibitem[{{Einasto} {et~al.}(2012{\natexlab{b}}){Einasto}, {Vennik}, {Nurmi},
  {Tempel}, {Ahvensalmi}, {Tago}, {Liivam{\"a}gi}, {Saar}, {Hein{\"a}m{\"a}ki},
  {Einasto}, \& {Mart{\'{\i}}nez}}]{Einasto2012}
{Einasto}, M., {Vennik}, J., {Nurmi}, P., {et~al.} 2012{\natexlab{b}}, \aap,
  540, A123

\bibitem[{{Fakhouri} \& {Ma}(2009)}]{Fakhouri2009}
{Fakhouri}, O. \& {Ma}, C.-P. 2009, \mnras, 394, 1825

\bibitem[{{Fukugita} {et~al.}(2004){Fukugita}, {Nakamura}, {Turner},
  {Helmboldt}, \& {Nichol}}]{Fukugita2004}
{Fukugita}, M., {Nakamura}, O., {Turner}, E.~L., {Helmboldt}, J., \& {Nichol},
  R.~C. 2004, \apjl, 601, L127

\bibitem[{{G{\'o}mez} {et~al.}(2003){G{\'o}mez}, {Nichol}, {Miller}, {Balogh},
  {Goto}, {Zabludoff}, {Romer}, {Bernardi}, {Sheth}, {Hopkins}, {Castander},
  {Connolly}, {Schneider}, {Brinkmann}, {Lamb}, {SubbaRao}, \&
  {York}}]{Gomez2003}
{G{\'o}mez}, P.~L., {Nichol}, R.~C., {Miller}, C.~J., {et~al.} 2003, \apj, 584,
  210

\bibitem[{{Gottl{\"o}ber} {et~al.}(2003){Gottl{\"o}ber}, {{\L}okas}, {Klypin},
  \& {Hoffman}}]{Gottlöber2003}
{Gottl{\"o}ber}, S., {{\L}okas}, E.~L., {Klypin}, A., \& {Hoffman}, Y. 2003,
  \mnras, 344, 715

\bibitem[{{Gunn} \& {Gott}(1972)}]{Gunn1972}
{Gunn}, J.~E. \& {Gott}, III, J.~R. 1972, \apj, 176, 1

\bibitem[{{Haas} {et~al.}(2011){Haas}, {Schaye}, \& {Jeeson-Daniel}}]{Haas2011}
{Haas}, M.~R., {Schaye}, J., \& {Jeeson-Daniel}, A. 2011, \mnras, 1812

\bibitem[{{Hamilton}(1988)}]{Hamilton1988}
{Hamilton}, A.~J.~S. 1988, \apjl, 331, L59

\bibitem[{{Hansen} {et~al.}(2009){Hansen}, {Sheldon}, {Wechsler}, \&
  {Koester}}]{Hansen2009}
{Hansen}, S.~M., {Sheldon}, E.~S., {Wechsler}, R.~H., \& {Koester}, B.~P. 2009,
  \apj, 699, 1333

\bibitem[{{Hickox} {et~al.}(2009){Hickox}, {Jones}, {Forman}, {Murray},
  {Kochanek}, {Eisenstein}, {Jannuzi}, {Dey}, {Brown}, {Stern}, {Eisenhardt},
  {Gorjian}, {Brodwin}, {Narayan}, {Cool}, {Kenter}, {Caldwell}, \&
  {Anderson}}]{Hickox2009}
{Hickox}, R.~C., {Jones}, C., {Forman}, W.~R., {et~al.} 2009, \apj, 696, 891

\bibitem[{{Hopkins} {et~al.}(2008){Hopkins}, {Hernquist}, {Cox}, \& {Kere{\v
  s}}}]{Hopkins2008}
{Hopkins}, P.~F., {Hernquist}, L., {Cox}, T.~J., \& {Kere{\v s}}, D. 2008,
  \apjs, 175, 356

\bibitem[{{Huertas-Company} {et~al.}(2011){Huertas-Company}, {Aguerri},
  {Bernardi}, {Mei}, \& {S{\'a}nchez Almeida}}]{Huertas2011}
{Huertas-Company}, M., {Aguerri}, J.~A.~L., {Bernardi}, M., {Mei}, S., \&
  {S{\'a}nchez Almeida}, J. 2011, \aap, 525, A157

\bibitem[{{Kauffmann} {et~al.}(2003){Kauffmann}, {Heckman}, {Tremonti},
  {Brinchmann}, {Charlot}, {White}, {Ridgway}, {Brinkmann}, {Fukugita}, {Hall},
  {Ivezi{\'c}}, {Richards}, \& {Schneider}}]{Kauffmann2003}
{Kauffmann}, G., {Heckman}, T.~M., {Tremonti}, C., {et~al.} 2003, \mnras, 346,
  1055

\bibitem[{{Kauffmann} {et~al.}(2004){Kauffmann}, {White}, {Heckman},
  {M{\'e}nard}, {Brinchmann}, {Charlot}, {Tremonti}, \&
  {Brinkmann}}]{Kauffmann2004}
{Kauffmann}, G., {White}, S.~D.~M., {Heckman}, T.~M., {et~al.} 2004, \mnras,
  353, 713

\bibitem[{{Kere{\v s}} {et~al.}(2005){Kere{\v s}}, {Katz}, {Weinberg}, \&
  {Dav{\'e}}}]{Keres2005}
{Kere{\v s}}, D., {Katz}, N., {Weinberg}, D.~H., \& {Dav{\'e}}, R. 2005,
  \mnras, 363, 2

\bibitem[{{Kewley} {et~al.}(2001){Kewley}, {Dopita}, {Sutherland}, {Heisler},
  \& {Trevena}}]{Kewley2001}
{Kewley}, L.~J., {Dopita}, M.~A., {Sutherland}, R.~S., {Heisler}, C.~A., \&
  {Trevena}, J. 2001, \apj, 556, 121

\bibitem[{{Komatsu} {et~al.}(2011){Komatsu}, {Smith}, {Dunkley}, {Bennett},
  {Gold}, {Hinshaw}, {Jarosik}, {Larson}, {Nolta}, {Page}, {Spergel},
  {Halpern}, {Hill}, {Kogut}, {Limon}, {Meyer}, {Odegard}, {Tucker}, {Weiland},
  {Wollack}, \& {Wright}}]{Komatsu:11}
{Komatsu}, E., {Smith}, K.~M., {Dunkley}, J., {et~al.} 2011, \apjs, 192, 18

\bibitem[{{Koopmann} \& {Kenney}(2004)}]{Koopmann2004}
{Koopmann}, R.~A. \& {Kenney}, J.~D.~P. 2004, \apj, 613, 851

\bibitem[{{Larson} {et~al.}(1980){Larson}, {Tinsley}, \&
  {Caldwell}}]{Larson1980}
{Larson}, R.~B., {Tinsley}, B.~M., \& {Caldwell}, C.~N. 1980, \apj, 237, 692

\bibitem[{{Lietzen} {et~al.}(2011){Lietzen}, {Hein{\"a}m{\"a}ki}, {Nurmi},
  {Liivam{\"a}gi}, {Saar}, {Tago}, {Takalo}, \& {Einasto}}]{Lietzen2011}
{Lietzen}, H., {Hein{\"a}m{\"a}ki}, P., {Nurmi}, P., {et~al.} 2011, \aap, 535,
  A21

\bibitem[{{Lietzen} {et~al.}(2009){Lietzen}, {Hein{\"a}m{\"a}ki}, {Nurmi},
  {Tago}, {Saar}, {Liivam{\"a}gi}, {Tempel}, {Einasto}, {Einasto}, {Gramann},
  \& {Takalo}}]{Lietzen2009}
{Lietzen}, H., {Hein{\"a}m{\"a}ki}, P., {Nurmi}, P., {et~al.} 2009, \aap, 501,
  145

\bibitem[{{Liivam{\"a}gi} {et~al.}(2012){Liivam{\"a}gi}, {Tempel}, \&
  {Saar}}]{Liivamägi2012}
{Liivam{\"a}gi}, L.~J., {Tempel}, E., \& {Saar}, E. 2012, \aap, 539, A80

\bibitem[{{Martini} {et~al.}(2007){Martini}, {Mulchaey}, \&
  {Kelson}}]{Martini2007}
{Martini}, P., {Mulchaey}, J.~S., \& {Kelson}, D.~D. 2007, \apj, 664, 761

\bibitem[{{Moore} {et~al.}(1996){Moore}, {Katz}, {Lake}, {Dressler}, \&
  {Oemler}}]{Moore1996}
{Moore}, B., {Katz}, N., {Lake}, G., {Dressler}, A., \& {Oemler}, A. 1996,
  \nat, 379, 613

\bibitem[{{Muldrew} {et~al.}(2012){Muldrew}, {Croton}, {Skibba}, {Pearce},
  {Ann}, {Baldry}, {Brough}, {Choi}, {Conselice}, {Cowan}, {Gallazzi}, {Gray},
  {Gr{\"u}tzbauch}, {Li}, {Park}, {Pilipenko}, {Podgorzec}, {Robotham},
  {Wilman}, {Yang}, {Zhang}, \& {Zibetti}}]{Muldrew2012}
{Muldrew}, S.~I., {Croton}, D.~J., {Skibba}, R.~A., {et~al.} 2012, \mnras, 419,
  2670

\bibitem[{{Park} {et~al.}(2008){Park}, {Gott}, \& {Choi}}]{Park2008}
{Park}, C., {Gott}, III, J.~R., \& {Choi}, Y.-Y. 2008, \apj, 674, 784

\bibitem[{{Pasquali} {et~al.}(2009){Pasquali}, {van den Bosch}, {Mo}, {Yang},
  \& {Somerville}}]{Pasquali2009}
{Pasquali}, A., {van den Bosch}, F.~C., {Mo}, H.~J., {Yang}, X., \&
  {Somerville}, R. 2009, \mnras, 394, 38

\bibitem[{{Porter} {et~al.}(2008){Porter}, {Raychaudhury}, {Pimbblet}, \&
  {Drinkwater}}]{Porter2008}
{Porter}, S.~C., {Raychaudhury}, S., {Pimbblet}, K.~A., \& {Drinkwater}, M.~J.
  2008, \mnras, 388, 1152

\bibitem[{{Postman} \& {Geller}(1984)}]{Postman1984}
{Postman}, M. \& {Geller}, M.~J. 1984, \apj, 281, 95

\bibitem[{{Press} \& {Davis}(1982)}]{Press1982}
{Press}, W.~H. \& {Davis}, M. 1982, \apj, 259, 449

\bibitem[{{Scudder} {et~al.}(2012){Scudder}, {Ellison}, \&
  {Mendel}}]{Scudder2012}
{Scudder}, J.~M., {Ellison}, S.~L., \& {Mendel}, J.~T. 2012, \mnras, 423, 2690

\bibitem[{{Silverman} {et~al.}(2009){Silverman}, {Kova{\v c}}, {Knobel},
  {Lilly}, {Bolzonella}, {Lamareille}, {Mainieri}, {Brusa}, {Cappelluti},
  {Peng}, {Hasinger}, {Zamorani}, {Scodeggio}, {Contini}, {Carollo}, {Jahnke},
  {Kneib}, {Le Fevre}, {Bardelli}, {Bongiorno}, {Brunner}, {Caputi}, {Civano},
  {Comastri}, {Coppa}, {Cucciati}, {de la Torre}, {de Ravel}, {Elvis},
  {Finoguenov}, {Fiore}, {Franzetti}, {Garilli}, {Gilli}, {Griffiths},
  {Iovino}, {Kampczyk}, {Koekemoer}, {Le Borgne}, {Le Brun}, {Maier},
  {Mignoli}, {Pello}, {Perez Montero}, {Ricciardelli}, {Tanaka}, {Tasca},
  {Tresse}, {Vergani}, {Vignali}, {Zucca}, {Bottini}, {Cappi}, {Cassata},
  {Marinoni}, {McCracken}, {Memeo}, {Meneux}, {Oesch}, {Porciani}, \&
  {Salvato}}]{Silverman2009}
{Silverman}, J.~D., {Kova{\v c}}, K., {Knobel}, C., {et~al.} 2009, \apj, 695,
  171

\bibitem[{{Skibba}(2009)}]{Skibba2009b}
{Skibba}, R.~A. 2009, \mnras, 392, 1467

\bibitem[{{Skibba} {et~al.}(2009){Skibba}, {Bamford}, {Nichol}, {Lintott},
  {Andreescu}, {Edmondson}, {Murray}, {Raddick}, {Schawinski}, {Slosar},
  {Szalay}, {Thomas}, \& {Vandenberg}}]{Skibba2009}
{Skibba}, R.~A., {Bamford}, S.~P., {Nichol}, R.~C., {et~al.} 2009, \mnras, 399,
  966

\bibitem[{{Sol Alonso} {et~al.}(2006){Sol Alonso}, {Lambas}, {Tissera}, \&
  {Coldwell}}]{SolAlonso2006}
{Sol Alonso}, M., {Lambas}, D.~G., {Tissera}, P., \& {Coldwell}, G. 2006,
  \mnras, 367, 1029

\bibitem[{{Sorrentino} {et~al.}(2006){Sorrentino}, {Antonuccio-Delogu}, \&
  {Rifatto}}]{Sorrentino2006}
{Sorrentino}, G., {Antonuccio-Delogu}, V., \& {Rifatto}, A. 2006, \aap, 460,
  673

\bibitem[{{Tago} {et~al.}(2008){Tago}, {Einasto}, {Saar}, {Tempel}, {Einasto},
  {Vennik}, \& {M{\"u}ller}}]{Tago2008}
{Tago}, E., {Einasto}, J., {Saar}, E., {et~al.} 2008, \aap, 479, 927

\bibitem[{{Tago} {et~al.}(2010){Tago}, {Saar}, {Tempel}, {Einasto}, {Einasto},
  {Nurmi}, \& {Hein{\"a}m{\"a}ki}}]{Tago2010}
{Tago}, E., {Saar}, E., {Tempel}, E., {et~al.} 2010, \aap, 514, A102

\bibitem[{{Tempel} {et~al.}(2009){Tempel}, {Einasto}, {Einasto}, {Saar}, \&
  {Tago}}]{Tempel2009}
{Tempel}, E., {Einasto}, J., {Einasto}, M., {Saar}, E., \& {Tago}, E. 2009,
  \aap, 495, 37

\bibitem[{{Tempel} {et~al.}(2011){Tempel}, {Saar}, {Liivam{\"a}gi}, {Tamm},
  {Einasto}, {Einasto}, \& {M{\"u}ller}}]{Tempel2011}
{Tempel}, E., {Saar}, E., {Liivam{\"a}gi}, L.~J., {et~al.} 2011, \aap, 529, A53

\bibitem[{{Tempel} {et~al.}(2012){Tempel}, {Tago}, \&
  {Liivam{\"a}gi}}]{Tempel:12}
{Tempel}, E., {Tago}, E., \& {Liivam{\"a}gi}, L.~J. 2012, \aap, 540, A106

\bibitem[{{Turner} \& {Gott}(1976)}]{Turner1976}
{Turner}, E.~L. \& {Gott}, III, J.~R. 1976, \apjs, 32, 409

\bibitem[{{van der Wel} {et~al.}(2010){van der Wel}, {Bell}, {Holden},
  {Skibba}, \& {Rix}}]{vanderWel2010}
{van der Wel}, A., {Bell}, E.~F., {Holden}, B.~P., {Skibba}, R.~A., \& {Rix},
  H.-W. 2010, \apj, 714, 1779

\bibitem[{{van Gorkom}(2004)}]{vanGorkom2004}
{van Gorkom}, J.~H. 2004, Clusters of Galaxies: Probes of Cosmological
  Structure and Galaxy Evolution, 305

\bibitem[{{Weinmann} {et~al.}(2006){Weinmann}, {van den Bosch}, {Yang}, \&
  {Mo}}]{Weinmann2006}
{Weinmann}, S.~M., {van den Bosch}, F.~C., {Yang}, X., \& {Mo}, H.~J. 2006,
  \mnras, 366, 2

\bibitem[{{Wilman} {et~al.}(2010){Wilman}, {Zibetti}, \&
  {Budav{\'a}ri}}]{Wilman2010}
{Wilman}, D.~J., {Zibetti}, S., \& {Budav{\'a}ri}, T. 2010, \mnras, 406, 1701

\bibitem[{{Wolf} {et~al.}(2009){Wolf}, {Arag{\'o}n-Salamanca}, {Balogh},
  {Barden}, {Bell}, {Gray}, {Peng}, {Bacon}, {Barazza}, {B{\"o}hm}, {Caldwell},
  {Gallazzi}, {H{\"a}u{\ss}ler}, {Heymans}, {Jahnke}, {Jogee}, {van Kampen},
  {Lane}, {McIntosh}, {Meisenheimer}, {Papovich}, {S{\'a}nchez}, {Taylor},
  {Wisotzki}, \& {Zheng}}]{Wolf2009}
{Wolf}, C., {Arag{\'o}n-Salamanca}, A., {Balogh}, M., {et~al.} 2009, \mnras,
  393, 1302

\bibitem[{{Yang} {et~al.}(2007){Yang}, {Mo}, {van den Bosch}, {Pasquali}, {Li},
  \& {Barden}}]{Yang2007}
{Yang}, X., {Mo}, H.~J., {van den Bosch}, F.~C., {et~al.} 2007, \apj, 671, 153

\bibitem[{{York} {et~al.}(2000){York}, {Adelman}, {Anderson}, {Anderson},
  {Annis}, {Bahcall}, {Bakken}, {Barkhouser}, {Bastian}, {Berman}, {Boroski},
  {Bracker}, {Briegel}, {Briggs}, {Brinkmann}, {Brunner}, {Burles}, {Carey},
  {Carr}, {Castander}, {Chen}, {Colestock}, {Connolly}, {Crocker}, {Csabai},
  {Czarapata}, {Davis}, {Doi}, {Dombeck}, {Eisenstein}, {Ellman}, {Elms},
  {Evans}, {Fan}, {Federwitz}, {Fiscelli}, {Friedman}, {Frieman}, {Fukugita},
  {Gillespie}, {Gunn}, {Gurbani}, {de Haas}, {Haldeman}, {Harris}, {Hayes},
  {Heckman}, {Hennessy}, {Hindsley}, {Holm}, {Holmgren}, {Huang}, {Hull},
  {Husby}, {Ichikawa}, {Ichikawa}, {Ivezi{\'c}}, {Kent}, {Kim}, {Kinney},
  {Klaene}, {Kleinman}, {Kleinman}, {Knapp}, {Korienek}, {Kron}, {Kunszt},
  {Lamb}, {Lee}, {Leger}, {Limmongkol}, {Lindenmeyer}, {Long}, {Loomis},
  {Loveday}, {Lucinio}, {Lupton}, {MacKinnon}, {Mannery}, {Mantsch}, {Margon},
  {McGehee}, {McKay}, {Meiksin}, {Merelli}, {Monet}, {Munn}, {Narayanan},
  {Nash}, {Neilsen}, {Neswold}, {Newberg}, {Nichol}, {Nicinski}, {Nonino},
  {Okada}, {Okamura}, {Ostriker}, {Owen}, {Pauls}, {Peoples}, {Peterson},
  {Petravick}, {Pier}, {Pope}, {Pordes}, {Prosapio}, {Rechenmacher}, {Quinn},
  {Richards}, {Richmond}, {Rivetta}, {Rockosi}, {Ruthmansdorfer}, {Sandford},
  {Schlegel}, {Schneider}, {Sekiguchi}, {Sergey}, {Shimasaku}, {Siegmund},
  {Smee}, {Smith}, {Snedden}, {Stone}, {Stoughton}, {Strauss}, {Stubbs},
  {SubbaRao}, {Szalay}, {Szapudi}, {Szokoly}, {Thakar}, {Tremonti}, {Tucker},
  {Uomoto}, {Vanden Berk}, {Vogeley}, {Waddell}, {Wang}, {Watanabe},
  {Weinberg}, {Yanny}, {Yasuda}, \& {SDSS Collaboration}}]{York:00}
{York}, D.~G., {Adelman}, J., {Anderson}, Jr., J.~E., {et~al.} 2000, \aj, 120,
  1579

\bibitem[{{Zandivarez} \& {Mart{\'{\i}}nez}(2011)}]{Zandivarez2011}
{Zandivarez}, A. \& {Mart{\'{\i}}nez}, H.~J. 2011, \mnras, 415, 2553

\end{thebibliography}
\end{document}